# DECAY OF KADOMTSEV–PETVIASHVILI LUMPS IN DISSIPATIVE MEDIA

**S. Clarke[1], K. Gorshkov[2], R. Grimshaw[3], and Y. Stepanyants[4,5*)]**

[1]School of Mathematical Sciences, Monash University, Clayton, Victoria 3800, Australia. E-mail: Simon.Clarke@monash.edu;

[2]Institute of Applied Physics of the Russian Academy of Sciences, Nizhny Novgorod, Russia. E-mail: Gorshkov@hydro.appl.sci-nnov.ru;

[3]University College London, UK. E-mail: R.Grimshaw@ucl.ac.uk;

[4]Nizhny Novgorod State Technical University n.a. R.E. Alekseev, Nizhny Novgorod, Russia;

[5]University of Southern Queensland, Toowoomba, QLD, 4350, Australia. E-mail: Yury.Stepanyants@usq.edu.au.

**Abstract**

The decay of Kadomtsev–Petviashvili lumps is considered for a few typical dissipations – Rayleigh dissipation, Reynolds dissipation, Landau damping, Chezy bottom friction, viscous dissipation in the laminar boundary layer, and radiative losses caused by large-scale dispersion. It is shown that the straight-line motion of lumps is unstable under the influence of dissipation. The lump trajectories are calculated for two most typical models of dissipation – the Rayleigh and Reynolds dissipations. A comparison of analytical results obtained within the framework of asymptotic theory with the direct numerical calculations of the Kadomtsev–Petviashvili equation is presented. Good agreement between the theoretical and numerical results is obtained.

___________________________________________________

*) Corresponding author: Yury.Stepanyants@usq.edu.au.

## I. Introduction

It is well-known that there are two versions of the Kadomtsev–Petviashvili (KP) equation describing quasi-plane waves in dispersive media. One of these versions, known as the KP2 equation, appears more frequently in many physical applications:

$$\frac{\partial}{\partial x}\left(\frac{\partial \upsilon}{\partial t} + c\frac{\partial \upsilon}{\partial x} + \alpha \upsilon \frac{\partial \upsilon}{\partial x} + \beta \frac{\partial^3 \upsilon}{\partial x^3}\right) = -\frac{c}{2}\frac{\partial^2 \upsilon}{\partial y^2}, \qquad (1)$$

where the coefficients $c > 0$, $\alpha$ (which can be of either sign) and $\beta$ ($\beta > 0$) depend on the environmental parameters of the particular medium (e.g., in application to water waves they depend on the depth, stratification, shear flow, etc). This equation, which can be considered as the two-dimensional generalisation of the classical Korteweg–de Vries (KdV) equation, has been studied extensively [see, for instance, [1, 2]].

There is also another version of the KP equation, the KP1 equation, with $\beta < 0$, which is also applicable to the description of weakly nonlinear waves in plasma, thin liquid films, thin plates, and solids [3–5]. In recent years numerous examples of the applicability of the KP1 equation have appeared in the scientific literature, including the description of nonlinear internal waves in shallow basins under the influence of Earth' rotation and shear flows [6].

A specific feature of the KP1 equation is the existence of lump solutions – solitary waves fully localised in space and algebraically decaying at infinity. These type of solitary waves are rather unusual, they preserve not only the shapes after interaction with each other, but do not experience even a phase shift, in contrast to exponentially decaying solitons (for details see [1, 2]). As well as single lumps, in the KP1 equation there is also a family of plane solitons and multi-lump solutions representing stationary moving fully localised formations with multi-humps [7, 26–28]. However all such solutions are unstable with respect to small perturbation [7, 26, 27], whereas single lumps are stable [29].

In real physical media usually there are different dissipative mechanisms which affect wave shapes and soliton dynamics. The influence of various types of dissipation on solitons and kinks has been studied in details for various evolution equations (see, for instance, [8, 9]). However, the influence of dissipation on the decay character of KP lumps has not been studied so far. In this paper we fill this gap and study the adiabatic decay of lumps under the effect of weak dissipations of various types; Rayleigh and Reynolds dissipation, as well as the dissipation caused by radiation of small-amplitude waves in the media containing weak low-frequency dispersion. The specific type of dissipation depends on the particular medium which can support the existence of KP lumps. For example, in plasma physics the most typical can be either

Rayleigh or Reynolds dissipation, or Landau damping [3, 5, 8, 22, 30]; the same (except the Landau damping) is true for waves on thin plates [4], in the case of internal waves in a rotating fluid with shear flows the most typical dissipation may be caused by large scale Coriolis dispersion [6, 18–20]; for surface waves on thin films the dissipation can be caused either by the bottom roughness or influence of a laminar bottom boundary layer [10, 16]. In all these cases lump decay can be studied from a common position by means of the adiabatic theory. Below we present the specific decay laws for all these cases, and in the last section we show that the lump trajectories on *x,y*-plane also depend on the type of dissipation.

**II. Adiabatic theory**

The KP1 equation (1) with $\beta < 0$ has a lump solution [1, 2]:

$$\upsilon(\xi, y, t) = -24\frac{V\beta}{\alpha} \frac{-3\beta + \frac{2}{c}V^2 y^2 + V(\xi - Vt)^2}{\left[-3\beta + \frac{2}{c}V^2 y^2 - V(\xi - Vt)^2\right]^2}, \qquad (2)$$

where $\xi = x - c\,t$, and $V < 0$ is the speed of a lump in the Galilean coordinate frame moving with the speed $c$ with respect to immovable observer. The amplitude of a lump is $A = 8V/\alpha$, and its total "mass" $M \equiv \int_{-\infty}^{+\infty}\int_{-\infty}^{+\infty} \upsilon(x, y, t)\,dx\,dy = 0$.

If weak dissipation is taken into account, then Eq. (1) is augmented by small additional terms whose structure depends on the nature of dissipation. In general such equation can be presented in the form

$$\frac{\partial}{\partial x}\left(\frac{\partial \upsilon}{\partial t} + c\frac{\partial \upsilon}{\partial x} + \alpha \upsilon \frac{\partial \upsilon}{\partial x} + \beta \frac{\partial^3 \upsilon}{\partial x^3} + \delta D[\upsilon]\right) = -\frac{c}{2}\frac{\partial^2 \upsilon}{\partial y^2}, \qquad (3)$$

where $D[\upsilon]$ is an operator which can be often expressed in the rather general form [10]:

$$D[\upsilon] = \frac{1}{\sqrt{2\pi}}\int_{-\infty}^{+\infty}(-ik)^m \tilde{\upsilon}(k,t)e^{ikx}dk. \qquad (4)$$

Here $\tilde{\upsilon}(k,t) = \frac{1}{\sqrt{2\pi}}\int_{-\infty}^{+\infty}\upsilon(x,t)e^{-ikx}dx$ is the Fourier transform of function $\upsilon(x,t)$, and the parameter $m$ depends on the specific type of dissipation. In particular, $m = 0$ (with $\delta > 0$) corresponds to linear Rayleigh damping when the dissipative term $\delta D(\upsilon)$ in Eq. (3) reduces simply to $\delta \upsilon$. Another widely used model with $m = 2$ corresponds to Reynolds dissipation (with $\delta < 0$); in the one-dimensional case Eq. (3) reduces to the KdV–Burgers equation, since then $D(\upsilon) = \upsilon_{xx}$. For modelling dissipation in a laminar bottom boundary layer the parameter $m = 1/2$ is customarily used. In some case the operator $D(\upsilon)$ is non-linear, for example, in the description of shallow water waves over a rugged bottom $D(\upsilon) = |\upsilon|\upsilon$ (a similar model is used sometimes for the description of dissipation in a turbulent bottom boundary layer).

When the dissipation is sufficiently small, one can expect that the basic structure of a lump remains the same, but its primary parameter $V$, which determines lump speed, amplitude and width, is no longer constant, but is a slow function of time. This dependence on time can be calculated by means of asymptotic theory [10–12], which presumes that the lump adiabatically varies with time, while retaining its shape. The asymptotic theory in essence reduces to the energy balance equation which eventually describes time dependence of the governing soliton or lump parameter.

Below we consider first the decay laws of ordinary symmetrical lumps as described by Eq. (2) under the action of different decay mechanisms. Then we study the dynamics of skew lumps in the $x,y$-plane for a few typical dissipations. We derive the trajectories of lumps and analyse them in detail.

**II.1. Rayleigh dissipation**

Consider first the effect of weak Rayleigh dissipation on the dynamics of a lump. The KP1 equation (1) with the Rayleigh dissipation can be presented in the form

$$\frac{\partial \upsilon}{\partial t} + c\frac{\partial \upsilon}{\partial x} + \alpha\upsilon\frac{\partial \upsilon}{\partial x} + \beta\frac{\partial^3 \upsilon}{\partial x^3} + \delta\upsilon = -\frac{c}{2}\frac{\partial^2}{\partial y^2}\int_{-\infty}^{x} \upsilon dx, \quad (5)$$

where $\delta > 0$ is the coefficient of dissipation.

Multiplying this equation by $\upsilon$ and integrating over $x$ and $y$ we obtain the energy balance equation:

$$\frac{1}{2}\frac{\partial}{\partial t}\langle \upsilon^2 \rangle = -\delta\langle \upsilon^2 \rangle, \quad \text{where} \quad \langle f(x,y) \rangle \equiv \int_{-\infty}^{+\infty}\int_{-\infty}^{+\infty} f(x,y)\,dxdy. \quad (6)$$

Note that this approach based on the energy balance equation agrees with the successive application of the asymptotic theory [10–12, 20], which presumes slow evolution of solitary wave parameters under the influence of a small perturbation (in our case the perturbation is caused by dissipation with a small coefficient $\delta \ll 1$).

Denoting the wave energy as $E = \langle\upsilon^2\rangle/2$, we see that it exponentially decays due to Rayleigh dissipation, $E = E_0 e^{-2\nu t}$, where $E_0$ is the initial lump energy. Substituting here the lump solution (2) and integrating over the infinite domain in the $x,y$-plane, we obtain for the lump energy (bearing in mind that $\beta < 0$ and $V < 0$):

$$E = -\frac{48\pi\beta}{\alpha^2}\sqrt{\frac{-cV}{2}}. \qquad (7)$$

Then, from Eq. (4) we find for the lump speed

$$\frac{dV}{dt} = -4\delta V. \qquad (8)$$

After integration we readily derive $V(t) = V_0 e^{-4\delta t}$ and similarly for the lump amplitude $A(t) = A_0 e^{-4\delta t}$. The derived dependence of lump amplitude on time can be compared with the similar dependence for the KdV soliton $A_{KdV}(t) = A_0 e^{-4\delta t/3}$ (see, e.g., [13] and references therein), whereas for linear waves $A_{lin}(t) = A_0 e^{-\delta t}$. Thus, the decay rate of a KP lump is three times greater than the decay rate of a KdV soliton and four times greater than the decay rate of a linear wave.

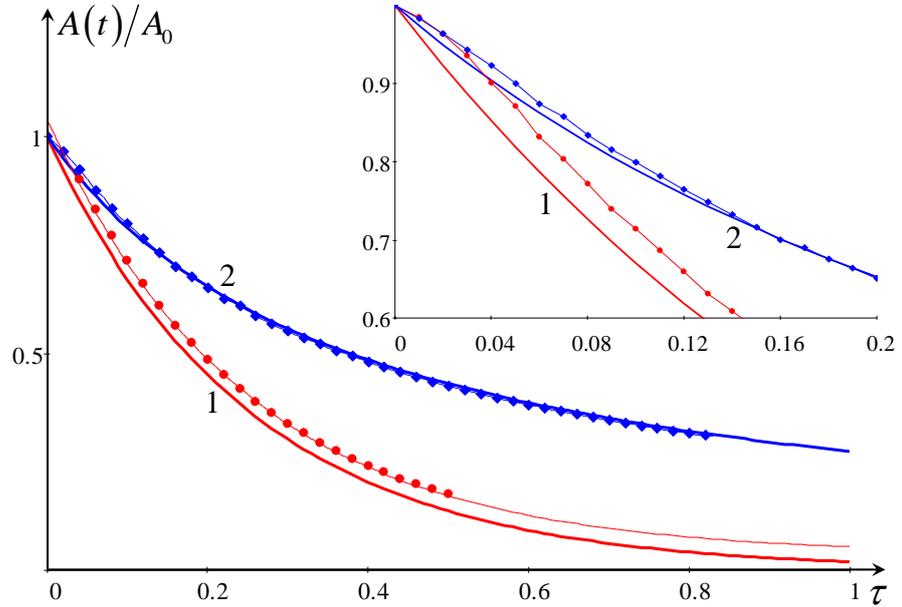

FIG. 1 (colour online). Lump amplitude dependence on normalised time $\tau = \delta t$ for different mechanisms of dissipation. Line 1 – Rayleigh dissipation; line 2 – Reynolds dissipation. Symbols represent numerical data with $\alpha = \beta = -1$, $c = 2$, and $\delta = 0.01$.

To validate this theoretical result we undertook a direct numerical simulation of lump evolution within the framework of the KP1 equation with the additional dissipative terms. The dependence of $A(t)$ for the Rayleigh dissipation as per adiabatic theory is shown in Fig. 1 by line 1, where $\tau = \delta t$. However, although the numerical data, which is shown by dots, follows the predicted decay rate, there is a noticeable shift of this data from the theoretical line. This can be explained by the influence of the initial non-adiabatic adjustment process, when a lump generates a small near-field perturbation (a "tail") and then decays adiabatically together with this tail (a similar phenomenon has been studied in one-dimensional case for the KdV equations with small dissipative terms [14]). The insert in Fig. 1 demonstrates the non-adiabatic decay of a lump at the initial stage of its evolution.

We do not consider here the formation and shape of the two-dimensional tail behind of the lump as this is a much more complicated problem than the counterpart in the one-dimensional case. However we illustrate the tail together with the lump in Fig. 2.

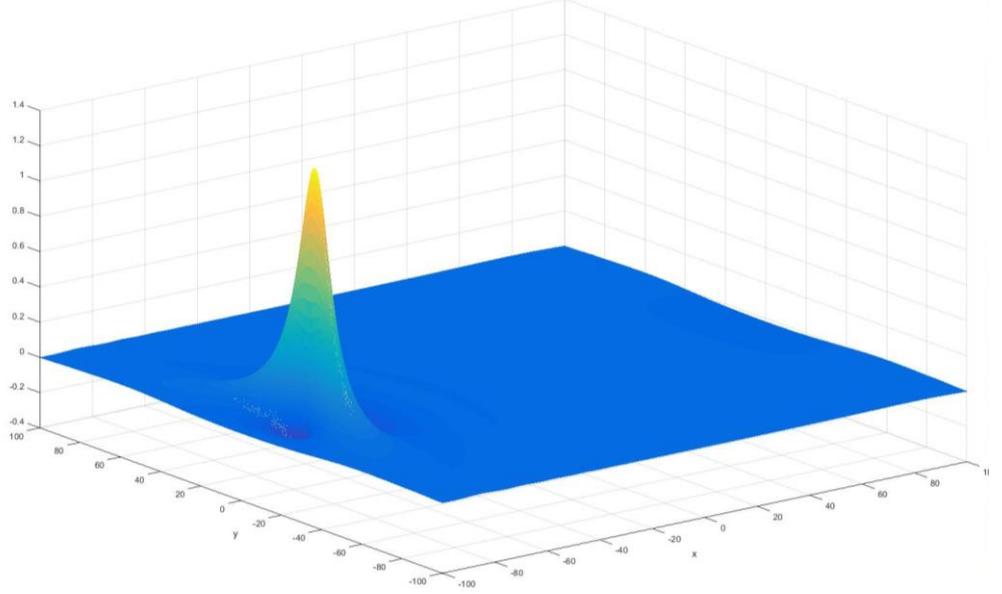

FIG. 2 (colour online). Lump profile at $t = 50$ within Eq. (5) with $\alpha = \beta = -1$, $c = 2$, and $\delta = 0.01$. The lump moves to the left being at the centre of the domain at $t = 0$.

### II.2. Reynolds dissipation

Consider now the effect of weak Reynolds dissipation on the dynamics of a KP lump. The KP1 equation with the Reynolds dissipation can be presented again in the form of Eq. (5) with the dissipative term $-\delta v_{xx}$ instead of $\delta v$, where indices of $v$ represent the derivatives with respect to $x$.

After multiplication of Eq. (5) by $v$ and integration over $x$ and $y$ we obtain now the following energy balance equation:

$$\frac{1}{2}\frac{\partial}{\partial t}\langle v^2 \rangle = -\delta \left\langle \left(\frac{\partial v}{\partial x}\right)^2 \right\rangle, \qquad (9)$$

where the angular brackets denote, as before, the integration over the entire $x,y$-plane.

As one can see, the wave energy $E = <\upsilon^2>/2$ now decays non-exponentially, but in a more complicated manner. Substituting in Eq. (9) the lump solution (2) and integrating over the infinite domain of the $x,y$-plane, we obtain for the lump speed

$$\frac{dV}{dt} = -\frac{8}{3}\frac{\delta}{\beta}V^2. \tag{10}$$

After integration of this equation we readily derive

$$V(t) = V_0\left(1 + \frac{8}{3}\frac{V_0}{\beta}\delta t\right)^{-1}; \quad A(t) = A_0\left(1 + \frac{\alpha A_0}{3\beta}\delta t\right)^{-1}. \tag{11}$$

The dependence $A(t)$ is depicted in Fig. 1 by line 2 together with the numerical data shown by rhombuses. Note that for this type of dissipation the agreement between the numerical data and adiabatic theory is almost perfect, except a very short transient period at the very beginning when the adjustment process occurs. In this case the numerical data quickly converges to the theoretical line 2. Figure 3 illustrates the tail behind the leading lump.

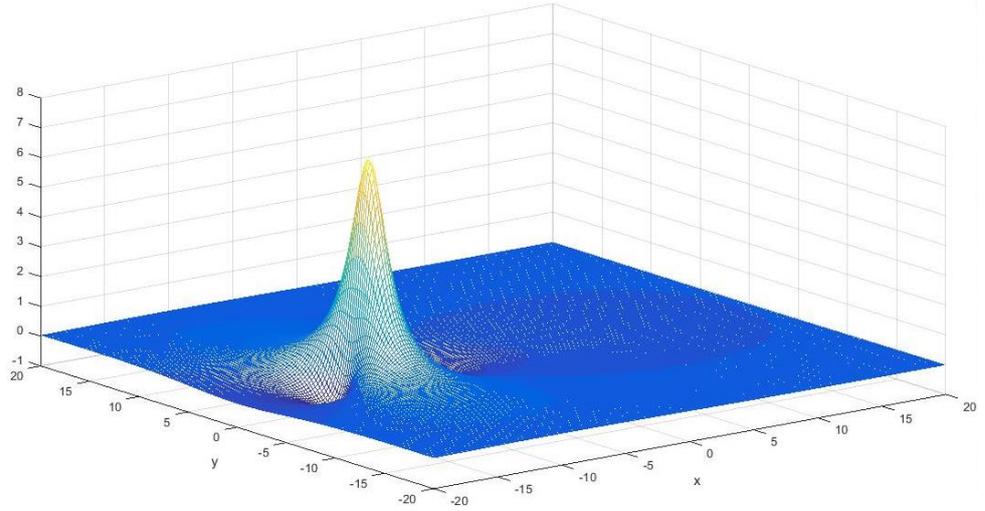

FIG. 3 (colour online). Lump profile at $t = 50$ within Eq. (5) with $\alpha = \beta = -1$, $c = 2$, and $\delta = 0.01$. The lump moves to the left being at the centre of the domain at $t = 0$.

Notice that the similar character of amplitude decay has been obtained for a KdV soliton, but with the characteristic decay time $T_{ch} = 45\beta/(4\alpha A_0 \delta)$ being 15/4 times greater than for a KP lump (see [15] and references therein). A small-amplitude sinusoidal wave decays exponentially with time with the decay rate $\delta k^2$, where $k$ is the wavenumber.

**II.3. Lump dissipation due to the influence of a laminar bottom boundary layer**

As has been mentioned above, in some cases the dissipative term in Eq. (3) can be nonlocal, but linear. In this case it can be expressed in terms of Eq. (4), where *m* is a non-even number. In particular, for modelling dissipation in a laminar bottom boundary layer the parameter $m = 1/2$ is customarily used. Here we consider this kind of dissipation, bearing in mind the possibility to observe lumps in weakly viscous thin layers of fluids, when the dissipation in a laminar bottom boundary layer can occur (notice that plane solitons of surface depression have been observed in a mercury [16]).

The energy balance equation determining lump soliton evolution in time in this case reads

$$\frac{1}{2}\frac{\partial}{\partial t}\langle \upsilon^2 \rangle = -\delta \langle \upsilon D[\upsilon] \rangle. \tag{12}$$

Using Parseval's theorem [17], the expressions in the angular brackets can be presented as

$$\langle \upsilon^2 \rangle = 2\int_{-\infty}^{+\infty}\left\{\frac{1}{\sqrt{2\pi}}\int_{-\infty}^{+\infty}|\tilde{\upsilon}|^2\,dk\right\}dy; \quad \langle \upsilon D[\upsilon] \rangle = \int_{-\infty}^{+\infty}\left\{\frac{1}{\sqrt{2\pi}}\int_{-\infty}^{+\infty}\text{Re}\left[(-ik)^m\right]|\tilde{\upsilon}|^2\,dk\right\}dy. \tag{13}$$

The Fourier spectrum of a lump can be easily evaluated

$$\tilde{\upsilon}(k,Y) = \frac{24\pi\beta}{\alpha}|k|e^{-|k|Y}, \quad \text{where} \quad Y^2 = \frac{3\beta}{V} - \frac{2V}{c}y^2. \tag{14}$$

Substituting this formula into Eq. (13) we obtain after evaluation of the integrals

$$\langle \upsilon^2 \rangle = -192\pi\frac{\beta}{\alpha^2}\sqrt{-\frac{cV}{2}}; \quad \langle \upsilon D[\upsilon] \rangle = \frac{N}{3\sqrt[4]{12}}\sqrt{\frac{c}{2}}\frac{(-V)^{3/4}}{(-\beta)^{5/4}}, \tag{15}$$

where $N = \int_0^\infty \text{sech}^{5/2}x \approx 0.874$.

Then from Eq. (12) we obtain

$$\frac{dV}{dt} = -\frac{\delta N \alpha^2 (-V)^{5/4}}{288\sqrt[4]{12}\pi(-\beta)^{9/4}}. \tag{16}$$

The solution to this equation is

$$V(t) = V_0\left(1 + \frac{t}{T_{ch}}\right)^{-4}, \tag{17}$$

where the characteristic decay time is $T_{ch} = \frac{1152\pi\beta^2}{N\delta\alpha^2}\sqrt[4]{\frac{12\beta}{V_0}}$ (the similar character of decay holds for a KdV soliton, see [10], for instance).

## II.4. Chezy dissipation

This type of dissipation, apparently, is not so topical for KP lumps. Nevertheless we will consider it in this section for the completeness. Note, however, that in the case of nonlinear waves on a thin liquid film over a rough surface this type of dissipation can be important. In general Chezy dissipation is used as an empirical model to describe wave energy losses due to scattering on random bottom roughness – see, e.g., the website "Bed roughness and friction factors in estuaries":

http://www.marinespecies.org/introduced/wiki/Bed_roughness_and_friction_factors_in_estuaries

In this case Eq. (5) contains the dissipation coefficient $\delta = \kappa|\upsilon|$ which is not a constant, but depends on the modulus of the wave field, and $\kappa$ is the coefficient of proportionality, which depends on the degree of roughness of the bottom [10]. The energy balance equation now reads:

$$\frac{1}{2}\frac{\partial}{\partial t}\langle \upsilon^2 \rangle = -\kappa \langle |\upsilon|\upsilon^2 \rangle. \tag{18}$$

Substituting here the lump solution (2), we obtain after simple manipulation

$$V(t) = V_0 \left(1 + 8\kappa V_0 t/\alpha \right)^{-1}; \quad A(t) = A_0 \left(1 + \kappa A_0 t\right)^{-1}. \tag{19}$$

The character of amplitude decay again is similar to that obtained earlier for a KdV soliton, but with the characteristic time of decay $T_{ch} = 15/(16\kappa A_0)$ being 15/16 times less than for a KP lump (see [10, 13] and references therein). It is also similar to the character of lump decay due to Reynolds dissipation (cf. Eq. (11)). However, the characteristic time of lump decay due to Chezy dissipation depends only on the initial amplitude and dissipation parameter $\kappa$, whereas in the case of Reynolds dissipation it depends also on the nonlinear and dispersion parameters.

Thus, one can see that KP lumps in general decay faster than KdV solitons in the media with the same mechanisms of dissipation. This can be explained by the two-dimensional structure of lumps – to preserve the lump shape in the two-dimensional space, lump wave energy should be distributed over a larger area than in the case of a plane soliton. Note that the total lump "mass" $M$ remains zero in the course of adiabatic decay, whereas the total "mass" of a KdV soliton decays proportional to the square root of its amplitude, $M(t) \equiv \int_{-\infty}^{+\infty} \upsilon(x,t)dx = \sqrt{A(t)}$ [8].

## II.5. Lump dissipation due to Landau damping

This type of lump decay may be essential in application to nonlinear waves in a magnetised plasma [3]. Decay of plane solitons within the framework of KdV equation augmented by the dissipative term describing the Landau damping was considered as earlier as 1969 [30]. In the case of two-dimensional waves described by the KP1 equation (3), the Landau damping corresponds to the integro-differential operator $D[\upsilon]$ (4) with the index $m = 1$ in the spectral form. It can be presented in the explicit form as

$$\frac{\partial}{\partial x}\left(\frac{\partial \upsilon}{\partial t} + c\frac{\partial \upsilon}{\partial x} + \alpha\upsilon\frac{\partial \upsilon}{\partial x} + \beta\frac{\partial^3 \upsilon}{\partial x^3} + \frac{\delta}{2\sqrt{2\pi}}\frac{\partial}{\partial x}\wp\int_{-\infty}^{+\infty}\frac{\upsilon(x',t)}{x-x'}dx'\right) = -\frac{c}{2}\frac{\partial^2 \upsilon}{\partial y^2}, \qquad (20)$$

where the symbol $\wp$ denotes the principal value of the integral.

The energy balance equation determining lump parameters evolution in time in this case reads

$$\frac{1}{2}\frac{\partial}{\partial t}\langle \upsilon^2 \rangle = -\frac{\delta}{2\sqrt{2\pi}}\left\langle \upsilon(x,t)\frac{\partial}{\partial x}\int_{-\infty}^{+\infty}\frac{\upsilon(x',t)}{x-x'}dx'\right\rangle. \qquad (21)$$

The expression in the right-hand side can be directly evaluated yielding

$$\left\langle \upsilon(x,t)\frac{\partial}{\partial x}\int_{-\infty}^{+\infty}\frac{\upsilon(x',t)}{x-x'}dx'\right\rangle = \frac{V^2}{144\beta^2}\sqrt{\frac{c}{-2V}}. \qquad (22)$$

Then, using Eq. (7) for $E$, we obtain

$$\frac{dV}{dt} = -\frac{\delta\alpha^2 V^2}{6912\sqrt{2\pi}\beta^3}. \qquad (23)$$

The solution to this equation is

$$V(t) = V_0\left(1 + \frac{t}{T_L}\right)^{-1}, \qquad (24)$$

where the characteristic decay time is $T_L = \dfrac{6912\sqrt{2\pi}\beta^3}{\delta\alpha^2 V_0}$. The similar character of decay holds for a KdV soliton (see [31]).

### III. Influence of large-scale dispersion on lump decay

In this section we consider the effect of large-scale dispersion on the dynamics of KP lumps. The KP1 equation augmented by the large-scale dispersion represents, in fact, the two-dimensional version of the Ostrovsky equation [18, 19, 9]:

$$\frac{\partial}{\partial x}\left(\frac{\partial \upsilon}{\partial t} + c\frac{\partial \upsilon}{\partial x} + \alpha\upsilon\frac{\partial \upsilon}{\partial x} + \beta\frac{\partial^3 \upsilon}{\partial x^3}\right) = -\frac{c}{2}\frac{\partial^2 \upsilon}{\partial y^2} + \gamma\upsilon, \qquad (25)$$

where $\gamma$ is the coefficient of large-scale dispersion (for details see the papers cited above). In the one-dimensional case the last term in Eq. (25) together with the dispersive term $\sim\beta$ provides the approximation of the dispersion relation in the intermediate range of wavenumbers: $\omega = ck - \beta k^3$ $+ \gamma/k$. This dispersion relation contains both the small scale-dispersion $\sim\beta$, (when $k \to \infty$) and large scale dispersion $\sim\gamma$ (when $k \to 0$).

For the derivation of the energy balance equation it is convenient to present Eq. (25) in the form of a set of two equations:

$$\begin{cases} \dfrac{\partial \upsilon}{\partial t} + c\dfrac{\partial \upsilon}{\partial x} + \alpha \upsilon \dfrac{\partial \upsilon}{\partial x} + \beta \dfrac{\partial^3 \upsilon}{\partial x^3} = w; \\ \dfrac{\partial w}{\partial x} = -\dfrac{c}{2}\dfrac{\partial^2 \upsilon}{\partial y^2} + \gamma \upsilon. \end{cases} \qquad (26)$$

After multiplication of the first equation of this set by $\upsilon$ and the second equation by $w$ and consequent integration over $x$ and $y$, we obtain the following set of equations:

$$\begin{cases} \dfrac{1}{2}\dfrac{d}{dt}\langle \upsilon^2 \rangle = \langle \upsilon w \rangle; \\ 0 = -\dfrac{c}{2}\left\langle w\dfrac{\partial^2 \upsilon}{\partial y^2}\right\rangle + \gamma \langle \upsilon w \rangle. \end{cases} \qquad (27)$$

Elimination of $\langle \upsilon w \rangle$ from this set yields

$$\dfrac{1}{2}\dfrac{d}{dt}\langle \upsilon^2 \rangle = \dfrac{c}{2\gamma}\left\langle w\dfrac{\partial^2 \upsilon}{\partial y^2}\right\rangle, \qquad w = \int\left(\gamma\upsilon - \dfrac{c}{2}\dfrac{\partial^2 \upsilon}{\partial y^2}\right)dx. \qquad (28)$$

After further manipulation and taking into account the zero lump mass ($M = 0$, Sect. II.4), we finally arrive to

$$\dfrac{1}{2}\dfrac{d}{dt}\langle \upsilon^2 \rangle = -\dfrac{c}{2}\int_{-\infty}^{+\infty}\left[\upsilon\dfrac{\partial}{\partial y}\left(\int \upsilon dx'\right)\right]\bigg|_{y=-\infty}^{y=+\infty} dx = 0. \qquad (29)$$

Thus, in this case the energy balance equation predicts that the soliton amplitude does not change under the action of large-scale dispersion. This contradicts with the data of direct numerical simulations shown below. However, the prediction (29) is based on the adiabatic assumption that the lump soliton keeps its shape under the influence of rotation. An asymptotic expansion of (26) based on the premise that $\gamma$ is small shows that this is not the case, and although such an expansion does confirm that the lump soliton energy is preserved at the leading order, examination of the $O(\gamma)$ correction term indicates that some trailing radiation is indeed generated due to the slow algebraic decay of the lump soliton on the far field, which would then lead to decay of the lump soliton at the next order in the asymptotic expansion. The precise determination of this effect is beyond the scope of the present article.

In Fig. 4 we present numerically obtained graphics of lump amplitude versus time for two cases, when the parameter $\gamma$ is negative (line 1) and positive (line 2); both cases are physically feasible. The numerical data for $\gamma < 0$ can be well-fitted by the empirical dependence

$$A(t) = A_0 \left(1 - \frac{t}{T_{ext}}\right)^2, \tag{30}$$

where the extinction time $T_{ext} = (C/\gamma)(\alpha A_0/12\beta)^{1/2}$, with $C = 3/4$. A similar formula was derived for the KdV soliton with $C = 1$ [19, 20].

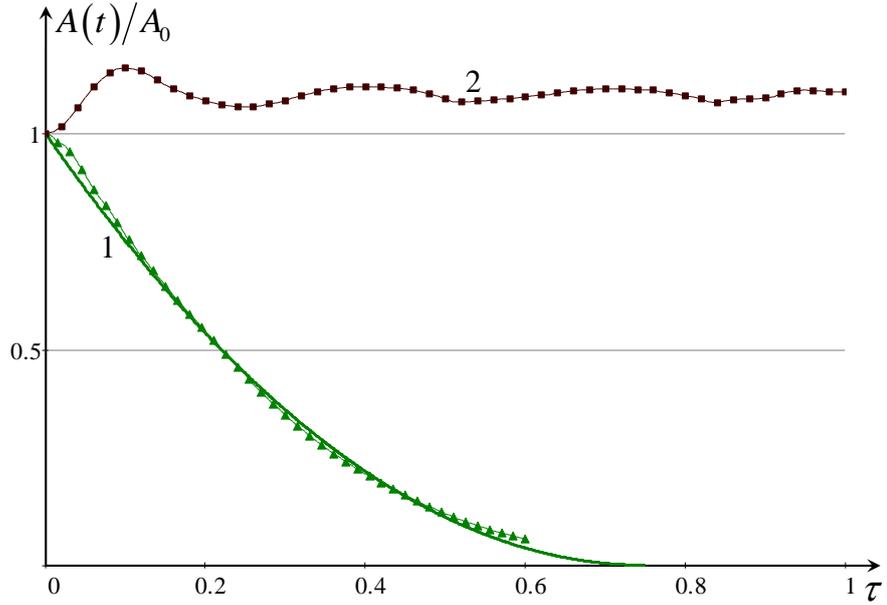

FIG. 4 (colour online). Lump amplitude dependence on normalised time $\tau = |\gamma|t$ under the influence of large-scale dispersion. Line 1 – $\gamma = -0.01$; line 2 – $\gamma = 0.01$; in both case we set $\alpha = \beta = -1$, $c = 2$. Symbols represent numerical data.

As one can see from Fig. 4, the numerical data declines from the empirical line 1 at the early transient stage. This is, apparently, can be explained again by the adjustment process when the lump radiates a tail and forms an adiabatically varying solitary wave. As mentioned above, we do not consider here the formation and shape of the two-dimensional tail behind the lump, but present only a few plots in Fig. 5 illustrating tale formation in the course of lump decay.

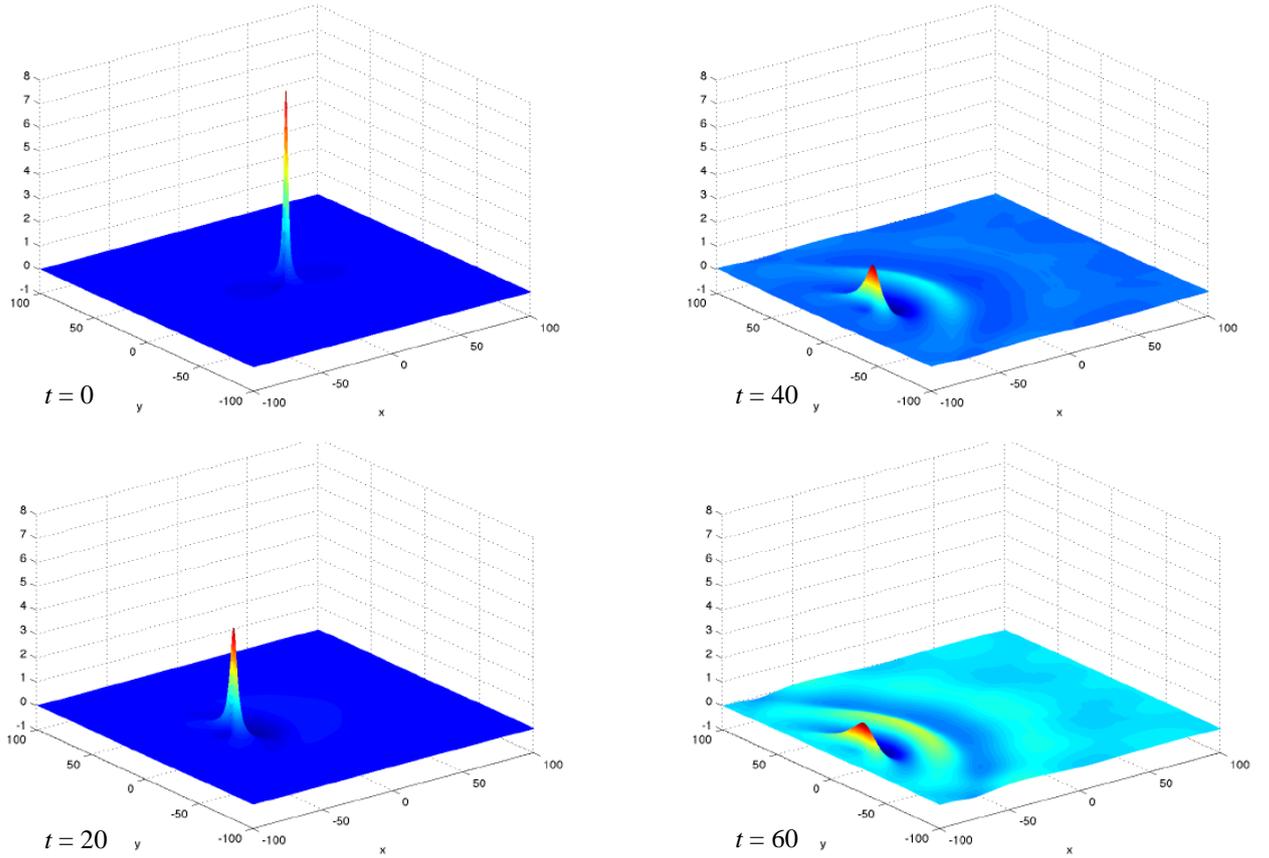

FIG. 5 (colour online). Lump profile at several instants of time shown in the each frame within Eq. (20) with $\alpha = \beta = -1$, $c = 2$, and $\gamma = 0.01$. The lump moves to the left being at the centre of the domain at $t = 0$.

In the case of negative $\gamma$ the initial lump does not decay, but gradually evolves to another lump of a higher amplitude (see the numerical data shown by square symbols in Fig. 4). A similar situation occurs with the KdV solitons which gradually transform into the Ostrovsky solitons [21] under the influence of large-scale dispersion. In Fig. 6 we present a stationary lump numerically obtained using Petviashvili's method [22, 23]. Under the influence of large-scale dispersion the lump becomes taller and narrower in comparison with the classical KP lump (2).

**IV. Skew lump dynamics on $x,y$-plane under the influence of dissipation**

As is well known, KP lumps can move under certain angles to the axis $x$ with the velocity $\mathbf{V} = (V_x, V_y)$ [1, 24, 25, 27, 28]. The corresponding formula for a skew lump moving obliquely in the $x,y$-plane without dissipation reads:

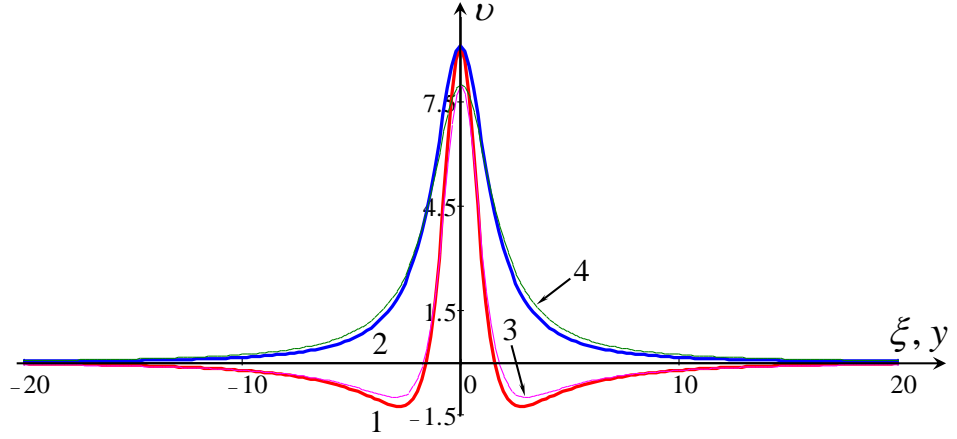

FIG. 6 (colour online). Cross-sections of stationary lump solution of the KP–Ostrovsky equation (20) for the following parameters: $\alpha = \beta = -1$, $c = 2$ and $\gamma = -0.01$. Line 1 – $\upsilon(\xi, 0)$, line 2 – $\upsilon(0, y)$. Thin lines 3 and 4 depict the corresponding cross-sections of the KP lump with the same parameters $\alpha$, $\beta$, $c$, and $\gamma = 0$.

$$\upsilon(\xi, y, t) = \frac{2V_x}{\alpha}(4-r^2)\frac{1+\frac{4-r^2}{12}\left[\frac{4-r^2}{4}Y^2 - \left(X + \frac{r}{2}Y\right)^2\right]}{\left\{1+\frac{4-r^2}{12}\left[\frac{4-r^2}{4}Y^2 + \left(X + \frac{r}{2}Y\right)^2\right]\right\}^2}, \qquad (31)$$

where $\xi = x - ct$, $V_x < 0$, and

$$X = (\xi - V_x t)\sqrt{\frac{V_x}{\beta}}; \quad Y = (y - V_y t)V_x\sqrt{\frac{-2}{\beta c}}; \quad r = V_y\sqrt{\frac{-2}{cV_x}}. \qquad (32)$$

The *y*-component of the velocity vector can be of either sign. In particular, when $V_y = 0$, then solution (31) reduces to solution (2).

Equations describing lump dynamics in the two-dimensional case can be derived from the asymptotic method developed in [11, 12]. Below we consider the particular case when only two dissipative terms are taken into account in Eq. (3), one of them corresponding to Rayleigh viscosity, $\nu\upsilon$, and another – to Reynolds viscosity, $-\delta\upsilon_{xx}$. Then the equations of motion reads:

$$\frac{dP_x}{dt} = -2\nu P_x - 2\delta\left(\frac{3}{4\pi}\right)^2 P_x^3, \qquad (33)$$

$$\frac{dP_y}{dt} = -2\nu P_y, \qquad (34)$$

where $P_x$ and $P_y$ are the components of the total wave momentum of a lump [7, 28]:

$$P_x = \frac{1}{2}\iint_{x,y}\upsilon^2(x, y, t)\,dx\,dy = -\frac{24\pi\beta}{\alpha^2}\sqrt{-2cV_x - V_y^2}, \qquad (35)$$

$$P_y = 2\sqrt{2} P_x \frac{V_y}{c} = -\frac{48\sqrt{2}\pi\beta}{\alpha^2 c} V_y \sqrt{-2cV_x - V_y^2} \ . \tag{36}$$

Equations (33) and (34) can be readily solved:

$$P_x = P_{x0} \frac{\sqrt{R} e^{-2\nu t}}{\sqrt{1 - e^{-4\nu t} + R}}, \qquad R = \frac{\nu}{\delta P_{x0}^2} \left(\frac{4\pi}{3}\right)^2, \tag{37}$$

$$P_y = P_{y0} e^{-2\nu t} \ . \tag{38}$$

From hereon indices 0 pertain to the initial values of the corresponding quantities.

From these equations we can find then the equations for the trajectory of a lump centre:

$$V_x \equiv \frac{d\xi_c}{dt} = -\frac{3}{8\pi c} \left(\frac{\alpha^2}{24\pi\beta} P_x\right)^2 - \frac{c}{16}\left(\frac{P_y}{P_x}\right)^2$$

$$= -\frac{2\pi}{3c}\left(\frac{\alpha^2}{24\pi\beta}\right)^2 \frac{\nu}{\delta} \frac{e^{-4\nu t}}{1 - e^{-4\nu t} + R} - \frac{c}{16} \frac{P_{y0}^2}{P_{x0}^2} \frac{1 - e^{-4\nu t} + R}{R}, \tag{39}$$

$$V_y \equiv \frac{dy_c}{dt} = \frac{c}{2\sqrt{2}} \frac{P_y}{P_x} = \frac{c}{2\sqrt{2}} \frac{P_{y0}}{P_{x0}} \frac{\sqrt{1 - e^{-4\nu t} + R}}{\sqrt{R}} \ . \tag{40}$$

After integration we finally derive:

$$\xi_c = \xi_{c0} - \frac{\pi}{3\delta c}\left(\frac{\alpha^2}{24\pi\beta}\right)^2 \ln\left|\frac{1 - e^{-4\nu t} + R}{R}\right| - \frac{c}{16R} \frac{P_{y0}^2}{P_{x0}^2}\left[(1+R)t - \frac{1 - e^{-4\nu t}}{4\nu}\right], \tag{41}$$

$$y_c = y_{c0} + \frac{c}{2\nu\sqrt{2R}} \frac{P_{y0}}{P_{x0}} \left[\sqrt{R} - \sqrt{1 - e^{-4\nu t} + R} + \sqrt{R+1}\left(2\nu t + \ln\frac{\sqrt{1 - e^{-4\nu t} + R} + \sqrt{R+1}}{\sqrt{R} + \sqrt{R+1}}\right)\right]. \tag{42}$$

### IV.1. Rayleigh dissipation

In the particular case, $\delta = 0$, Eqs. (33), (34) and (37), (38) reduce to

$$\frac{dP_x}{dt} = -2\nu P_x, \qquad \frac{dP_y}{dt} = -2\nu P_y. \tag{43}$$

$$P_x = P_{x0} e^{-2\nu t}, \qquad P_y = P_{y0} e^{-2\nu t}. \tag{44}$$

The equations for the trajectory of a lump centre read:

$$V_x \equiv \frac{d\xi_c}{dt} = -\frac{1}{2c}\left(\frac{\alpha^2 P_{x0}}{24\pi\beta}\right)^2 e^{-4\nu t} - \frac{c}{16}\left(\frac{P_{y0}}{P_{x0}}\right)^2, \tag{45}$$

$$V_y \equiv \frac{dy_c}{dt} = \frac{c}{2\sqrt{2}} \frac{P_{y0}}{P_{x0}} \ . \tag{46}$$

As follows from these equations, a transverse component of a skew-lump velocity remains constant, whereas the longitudinal component consists of two terms; one of them exponentially decays in time, whereas the other one remains constant. Therefore when the first exponential term vanishes, the lump centre formally moves with a constant speed at some angle $\theta$ to the $x$-axis, where $\theta = -\text{atan}\,(4\sqrt{2}P_{x0}/P_{y0})$, but actually the lump amplitude is zero at this stage.

Solutions of equations (45), (46) are:

$$\xi_c = \xi_{c0} - \frac{3}{32\pi\nu c}\left(\frac{\alpha^2}{24\pi\beta}P_{x0}\right)^2 \left(1 - e^{-4\nu t}\right) - \frac{c}{16}\left(\frac{P_{y0}}{P_{x0}}\right)^2 t, \qquad (47)$$

$$y_c = y_{c0} + \frac{c}{2\sqrt{2}}\frac{P_{y0}}{P_{x0}}t. \qquad (48)$$

As follows from Eq. (47), the path traversed by the lump due to viscous decay is:

$$X_{ext} = -\frac{3}{32\pi\nu c}\left(\frac{\alpha^2}{24\pi\beta}P_{x0}\right)^2. \qquad (49)$$

An example of lamp trajectory is shown in Fig. 7 by line 1.

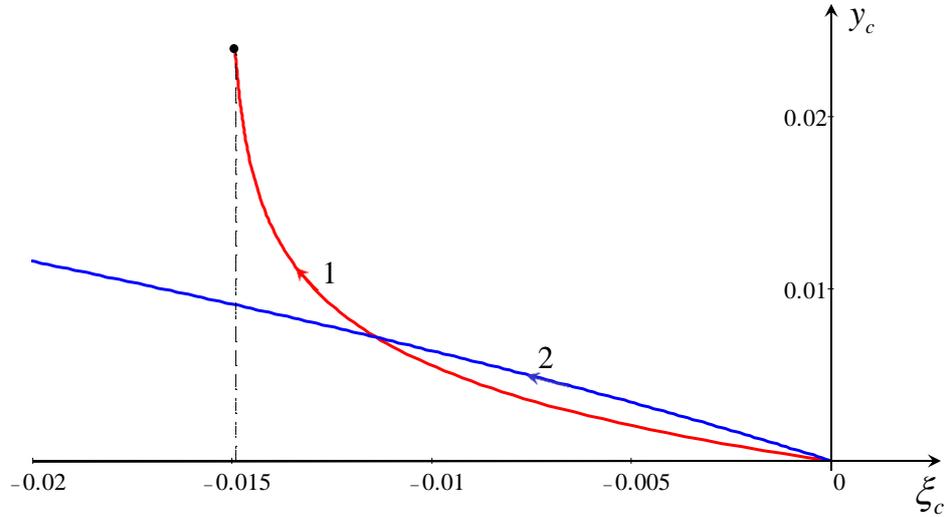

FIG. 7 (colour online). An example of lump trajectory in the medium with Rayleigh dissipation (line 1) and with Reynolds dissipation (line 2; the horizontal coordinate $\xi_c$ in this case is 100 times compressed). The vertical dashed line shows the extinction distance $X_{ext}$ at which a lump completely vanishes as it approaches the dot.

**IV.2. Reynolds dissipation**

In another particular case, $\nu = 0$, Eqs. (33), (34) and (37), (38) reduce to

$$\frac{dP_x}{dt} = -2\delta\left(\frac{3}{4\pi}\right)^2 P_x^3, \qquad \frac{dP_y}{dt} = 0. \qquad (50)$$

Interestingly, the *y*-component of the momentum does not change in time in this case. Solution to this system is:

$$P_x = \frac{P_{x0}}{\sqrt{1+t/T}}, \qquad P_y = P_{y0}, \tag{51}$$

where $T = (4\pi P_{x0}/3)^2/4\delta$.

The equations for the trajectory of a lump centre read:

$$V_x \equiv \frac{d\xi_c}{dt} = -\frac{1}{2c(1+t/T)}\left(\frac{\alpha^2 P_{x0}}{24\pi\beta}\right)^2 - \frac{c(1+t/T)}{16}\left(\frac{P_{y0}}{P_{x0}}\right)^2, \tag{52}$$

$$V_y \equiv \frac{dy_c}{dt} = \frac{c}{2\sqrt{2}}\frac{P_{y0}}{P_{x0}}\sqrt{1+\frac{t}{T}}. \tag{53}$$

In contrast with the previous case of Rayleigh dissipation, now both components of the lump velocity vary in time, and the trajectory of lump centre is described by the following equations:

$$\xi_c = \xi_{c0} - \frac{T}{2c}\left(\frac{\alpha^2 P_{x0}}{24\pi\beta}\right)^2 \ln\left(1+\frac{t}{T}\right) - \frac{cT}{32}\left(\frac{P_{y0}}{P_{x0}}\right)^2\left[\left(\frac{t}{T}+1\right)^2 - 1\right], \tag{54}$$

$$y_c = y_{c0} + \frac{cT}{3\sqrt{2}}\frac{P_{y0}}{P_{x0}}\left(1+\frac{t}{T}\right)^{\frac{3}{2}}. \tag{55}$$

An example of the lump trajectory for this case is shown in Fig. 7 by line 2.

**V. Conclusions**

Thus, in this paper we have calculated the decay of KP lumps under the influence of various mechanisms of dissipation. The dependences of lump amplitude and velocity on time were found within the framework of adiabatic theory for a few typical dissipations (Rayleigh dissipation, Reynolds dissipation, Landau damping, Chezy bottom friction, viscous dissipation in the laminar boundary layer, and radiative losses caused by large-scale dispersion). A comparison of analytical results with the direct numerical calculations within the KP1 equation demonstrates a good agreement between the theoretical results and numerical data. In the Table below we summarise the decay laws of KP lumps for different types of dissipation.

It has been shown that the straight-line motion of lumps in dissipative media is unstable. If a lump commences its motion under a small angle with respect to the *x*-axis, it declines further from the axis, as shown in Fig. 7. In the case of Rayleigh dissipation the lump tends to move eventually in the perpendicular direction, along the *y*-axis. In general, lump trajectories on the *x*,*y*-plane depend on the dissipation mechanism.

Table. Summary of decay laws of KP lumps at different types of dissipation

| Type of dissipation | Decay character | Characteristic decay time |
|---|---|---|
| Rayleigh dissipation | $A(t) = A_0 e^{-t/\tau}$ | $\tau = \dfrac{1}{4\delta}$ |
| Reynolds dissipation | $A(t) = \dfrac{A_0}{1 + t/\tau}$ | $\tau = \dfrac{3\beta}{\delta \alpha A_0}$ |
| Landau damping | $A(t) = \dfrac{A_0}{1 + t/\tau}$ | $\tau = 55296\sqrt{2\pi}\,\dfrac{\beta^3}{\delta \alpha^3 A_0}$ |
| Chezy dissipation | $A(t) = \dfrac{A_0}{1 + t/\tau}$ | $\tau = \dfrac{1}{\kappa A_0}$ |
| Dissipation in a bound. layer | $A(t) = \dfrac{A_0}{(1 + t/\tau)^4}$ | $\tau = 1.3 \cdot 10^4 \dfrac{\beta^2}{\delta \alpha^2} \sqrt[4]{\dfrac{\beta}{\alpha A_0}}$ |
| Radiative decay | $A(t) = A_0 \left(1 - \dfrac{t}{\tau}\right)^2$ | $\tau = \dfrac{1}{8\gamma}\sqrt{\dfrac{3\alpha A_0}{\beta}}$ |

In conclusion we note that the results of this study can be very relevant to the dynamics of internal wave lumps whose existence is supported by simultaneous influence of Earth' rotation and shear flows [6]. This is very interesting and topical problem for the coastal zone oceanography and worth further investigation.

**Acknowledgments.** R.G. was supported by the Leverhulme Trust through the award of a Leverhulme Emeritus Fellowship. Y.S. acknowledges the funding of this study from the State task program in the sphere of scientific activity of the Ministry of Education and Science of the Russian Federation (Project No. 5.1246.2017/4.6).